\documentclass[prl,aps,twocolumn,superscriptaddress]{revtex4}
\usepackage{graphicx,dcolumn}
\usepackage{amsfonts,amsmath,amssymb,bm,bbm, mathtools, siunitx}
\usepackage{float}
\usepackage{color}
\usepackage{physics}
\usepackage[toc,page]{appendix}

\def\be{\begin{eqnarray}}
\def\ee{\end{eqnarray}}
\def\ben{\begin{eqnarray*}}
\def\een{\end{eqnarray*}}
\def\bes{\begin{subequations}}
\def\ees{\end{subequations}}

\def\nn{\nonumber}

\newcommand{\wig}[1]{\mathrel{\hbox{\hbox to 0pt{\lower.6ex\hbox{$\sim$}\hss    }\raise.4ex\hbox{$#1$}}}}
\newcommand\sss{\scriptscriptstyle}

\begin{document}

\title{Nature of Non-Adiabatic Electron-Ion Forces in Liquid Metals}
%

\author{Jacopo \surname{Simoni}}
\email{jsimoni@lanl.gov}
\author{J\'er\^ome \surname{Daligault}}
\email{daligaul@lanl.gov}
\affiliation{Los Alamos National Laboratory, Los Alamos, NM 87545, USA}

\begin{abstract}
An accurate description of electron-ion interactions in materials is crucial for our understanding of their equilibrium and non-equilibrium properties.
Here, we assess the properties of frictional forces experienced by ions in non-crystalline metallic systems, including liquid metals and warm dense plasmas, that arise from electronic excitations driven by the nuclear motion due to the presence of a continuum of low-lying electronic states.
To this end, we perform detailed ab-initio calculations of the full friction tensor that characterizes the set of friction forces.
The nonadiabatic electron-ion interactions introduce hydrodynamic couplings between the ionic degrees of freedom, which are sizeable between nearest neigbors.
The friction tensor is generally inhomogeneous, anisotropic and non-diagonal, especially at lower densities.
\end{abstract}


\date{\today}

\maketitle

The study of interactions between electrons and ions in metals is a cornerstone of condensed matter physics.
Despite the small electron to ion mass ratio, the interactions are never strictly adiabatic as a continous spectrum of electronic excitations of arbitrarily low energy is available at the Fermi level to couple with the nuclear motions \cite{Dou2018,Li1992,Li1995}.
Similar couplings influence a host of physical and chemical processes at metal surfaces \cite{Dou2018,Wodtke2004,Tully2012}, which has generated a great deal of experimental and theoretical interest \cite{Askerka2016,Maurer2016,Rittmeyer_2016,Novko_2015,White2005,Nahler2008,Shenvi2009}.
The non-adiabatic transitions result in exchanges of small amounts of energy that maintain thermal equilibrium between electrons and ions, and drive the irreversible evolution towards thermal equilibrium from a non-equilibrium state \cite{Race2010,Tamm2018,Simoni_2019,Magyar2016,Nazarov2007,Nazarov2008}.
In solid metals, the non-adiabatic interactions are well understood in terms of electron-phonon interactions \cite{Giustino2017}.
In metallic systems where the electron-phonon picture no longer holds because ions have the ability to travel throughout the system, such as in liquid metals and in warm dense plasmas created in various matter under extreme conditions experiments, the basic properties of these nonadiabatic electron-ion interactions remain largely unexplored.


In general, a detailed description of nonadiabatic couplings with first principles simulations remains a formidable challenge.
Fortunately, a simplified coarse-grained description that avoids the explicit propagation of the electron dynamics is possible
\cite{Daligault_Mozyrsky_2018,Daligault_Mozyrsky_2009}.
Indeed, 1) the small electron to ion mass ratio -- or, more accurately, the large electron to ion velocity ratio -- and 2) the presence of a continuum of electronic states imply the existence of two distinct time scales $t_\gamma$ and $\tau_e$, respectively related to the slow relaxation of ionic momenta induced by electronic frictional forces and the fast electronic density fluctuations, such that the variations of the ion velocities over an interval of time $\delta t$ with $\tau_e\!\ll\!\delta t\!\ll\! t_\gamma$ satisfy the Langevin equations \cite{Daligault_Mozyrsky_2018}
\be
M\ddot{R}=-\nabla_R V_{ii}-\sum_n{p_n\frac{\partial E_n}{\partial R}}-M\/\tensor{\gamma}(R)\cdot\dot{R}+\xi(R) \,,
\label{generalized_Langevin}
\ee
while the electron dynamics is described by a master equation $\dot{p}_n\!=\!\sum_m{\left[W_{nm}p_m-W_{mn}p_n\right]}$
for the populations $p_n$ of the adiabatic electronic states $E_n(R)$ \cite{Daligault_Mozyrsky_2018}.
By way of illustration, for aluminum at $2.35$ $\rm g.cm^{-3}$ and $0.5$ eV, we find $\tau_e< 1$ $\rm fs$ and $t_\gamma\simeq 5$ $\rm ps$. 
In Eq.~(\ref{generalized_Langevin}), $R\!=\!\{{\bf R}_a\}\!=\!\{R_{ax}\}$ denotes the set of Cartesian positions of all the ions, $V_{ii}(R)$ is the interaction energy between ions,  $N$ and $M$ are the total number of ions and the ion mass, and $\tensor{\gamma}\!=\!\{\gamma_{ax,by}\}$ is the $3N\!\times\! 3N$-dimensional electronic friction tensor.
The second term in the rhs of Eq.(\ref{generalized_Langevin}) is the adiabatic force; in thermal equilibrium, it reduces to the usual Born-Oppenheimer (BO) force describing the interaction between the ions and the density the electrons would have if they were in thermal equilibrium with the instantaneous ionic configuration $R$.
The other force terms account for the fact that electrons do not adjust instantaneously to ionic motions.
The frictional forces $-\!M\gamma_{ax,by}(R)\dot{R}_{by}$ arise from the electronic excitations induced by ion $a$'s own motion ($b=a$), or by the motion of all other ions ($b\neq a$) and mediated to $a$ by the conduction electrons.
Finally, $\xi$ is a white-noise random force caused by the rapidly varying electronic density fluctuations.
In thermal equilibrium, $p_n$ is given by the normalized Boltzmann factor $e^{-E_n/k_{\sss B}T}/{\cal{Z}}$, and the frictional and random forces completely determine each other via a fluctuation-dissipation relation.
Each $\gamma_{ax,by} $ is related to the correlation function of the fluctuating electron-ion forces on $a$ and $b$ such as \cite{Daligault_Mozyrsky_2018,Daligault2019}
\be
\lefteqn{\gamma_{ax,by}(R)=\frac{1}{2Mk_{\sss B}T}\times}&&\label{gamma_ax_by}\\
&&\hspace{-0.25cm}{\rm Re}\int_0^\infty{\!\!\!\!\!dt\!\! \int\!\!\!\int{\!\!d{\bf r}d{\bf r}^{\sss\prime}\nabla_{ax}V_{ei}({\bf r}) \big\langle\!\delta \hat{n}_e({\bf r},t)\delta \hat{n}_e({\bf r}^{\sss\prime}, 0)\!\big\rangle_{\!\sss R}\!\nabla_{by}V_{ei}({\bf r}^{\sss\prime})}}\nn
\ee
where $V_{ei}({\bf r})\!=\!\sum_a{v_{ei}({\bf r}\!-\!{\bf R}_a)}$ is the total electron-ion potential energy and the correlation function $\big\langle\!\delta \hat{n}_e({\bf r},t)\delta \hat{n}_e({\bf r}^{\sss\prime}, 0)\!\big\rangle_{\!\sss R}$ describes the dynamics of the electron density fluctuations in the ionic configuration $R$.

In this work, we use first-principles simulations to measure the strength of electronic friction and assess the importance of its tensorial properties and of its dependence on the instantaneous ionic positions $R$.
Following the method discussed in Ref.\cite{Simoni_2019}, the friction coeffcients (\ref{gamma_ax_by}) are calculated from the electronic and ionic structures obtained with Density Functional Theory (DFT) based quantum molecular dynamics (QMD) simulations.
Full details on the simulations are given in the Supplemental Material (SM) \cite{SM}.
To help comprehend the data, we compare them with two limiting models.
Because liquid metals and plasmas are isotropic and homogeneous at large scales, the tensor $\tensor{\gamma}(R)$ greatly simplifies when averaged over a thermal ensemble.
Thus, the canonical average of the ``self'' terms ($a\!=\!b$) over all ionic configurations with $a$ fixed satisify $\left\langle \gamma_{ax,ay}(R)\right\rangle=\gamma_d\delta_{x,y}$ where $\gamma_d$ is independent of ${\bf R}_a$.
Similarly, the canonical average of the ``cross'' terms ($a\!\neq\! b$) over all configurations with $a$ and $b$ fixed are diagonal in the coordinate system where the $x$ axis is directed along the interparticle direction ${\bf R}_{ab}={\bf R}_a-{\bf R}_b$ with $\left\langle \gamma_{ax,by}(R)\right\rangle=\gamma^\parallel(R_{ab})\delta_{x,x}+\gamma^\perp(R_{ab}) \left[\delta_{y,y}+\delta_{z,z}\right]$.
In addition, we compare our results with the model
\be
\widetilde{\gamma}_{ax,by}=-\frac{1}{M}\!\int{\!\!\frac{d{\bf k}}{(2\pi)^3}\left|\frac{v_{ei}(k)}{\epsilon(k,0)}\right|^2\!\frac{\partial{\rm Im}\chi_0(k,0)}{\partial\omega} k_xk_ye^{i{\bf k}\cdot{\bf R}_{ab}}}\,, \label{tilde_gamma_axby}
\ee
obtained by approximating Eq.(\ref{gamma_ax_by}) to second order in the electron-ion potential or, equivalently, by
substituting in Eq.(\ref{gamma_ax_by}) the density correlation function of the homogeneous electron gas (jellium) model.
Here, $\epsilon(k,\omega)$ is the jellium dielectric function and $\chi_0(k,\omega)$ the ideal gas response function \cite{Daligault2019}.
At this order, $\tensor{\gamma}$ has the same symmetry properties as $\left\langle \tensor{\gamma}\right\rangle$, and the diagonal $\widetilde{\gamma}_d$ reduces to a celebrated model for the energy loss by slow ions in a jellium \cite{FerrellRitchie1977}.

\begin{figure}[t]
\includegraphics[scale=0.9]{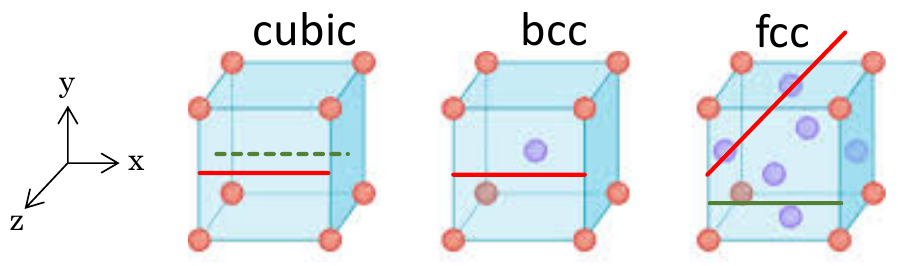}\\
\includegraphics[width=\columnwidth]{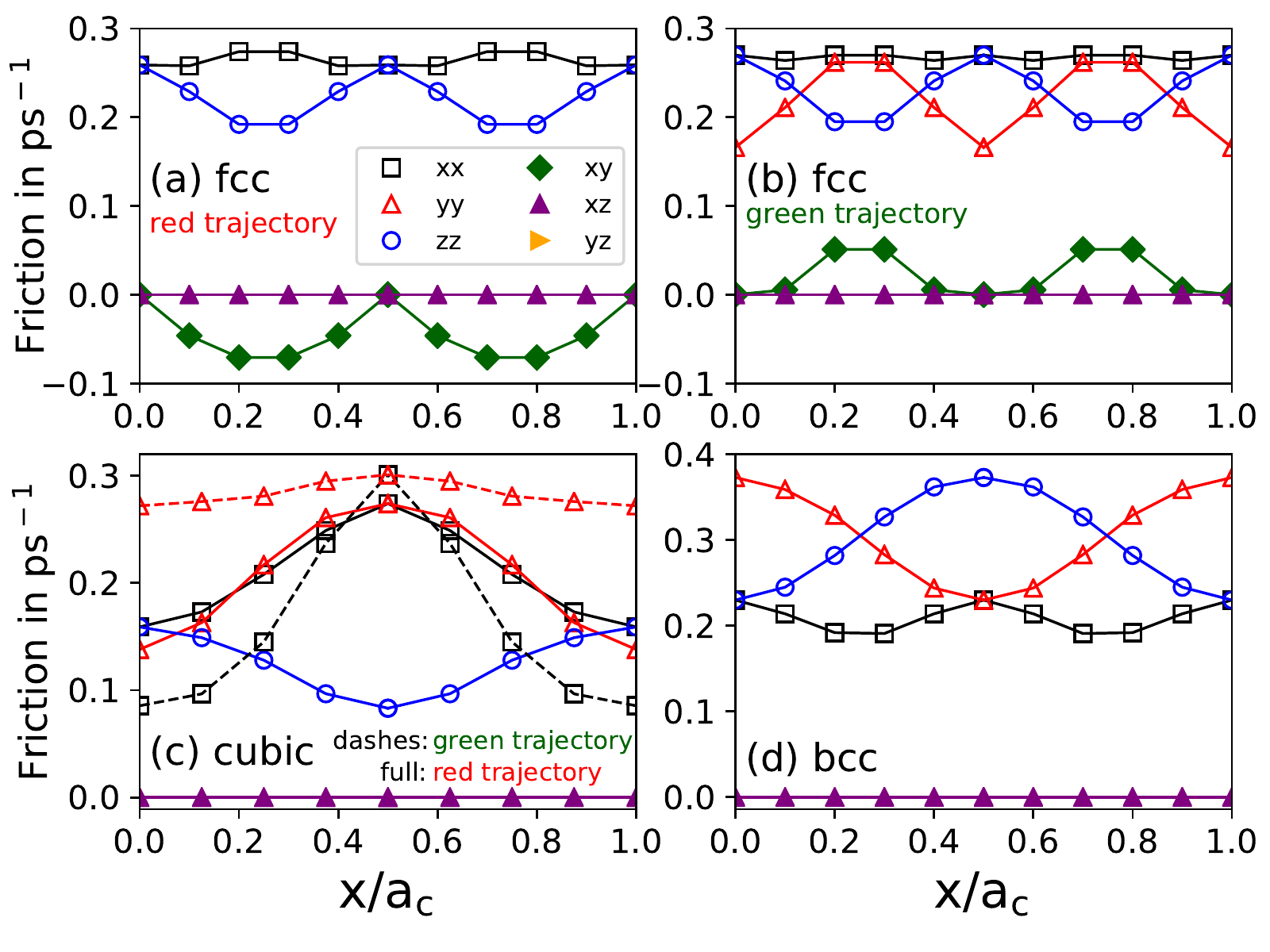}
\vspace{-0.85cm}
\caption{(Color online) Diagonal and off-diagonal friction coefficients felt by a proton immersed in perfect crystals of aluminum at $2.7$ $\rm g.cm^{-3}$ and $0.1$ $\rm eV$.
The proton is displaced along the red and green line segments shown in the cartoon (see details in \cite{SM}).
The horizontal axis is the proton coordinate along $x$ in units of the cell size $a_c$}
\label{Fig:2}
\end{figure}

We begin with simple illustrative calculations to familiarize oneself with the friction tensor.
Figure~\ref{Fig:2} shows the self components $\gamma_{px,py}$ ($\gamma_{xy}$ for short) felt by a proton $p$ immersed in perfect crystalline structures of Al at normal density $2.7$ $\rm g.cm^{-3}$.
Three stuctures are considered, namely FCC (panels a and b), simple cubic (c) and BCC (d).
The proton position ${\bf R}_p$ is varied in a unit cell along the rectilinear segments illustrated in the cartoon.
At each ${\bf R}_p$, the thermal electronic structure is calculated assuming an electronic temperature of $0.1$ $\rm eV$, and is used in Eq.(\ref{gamma_ax_by}) to calculate $\gamma_{xy}({\bf R}_p)$.
The results highlight important properties of the instantaneous friction tensor.
Firstly, each $\gamma_{xy}$ generally depends on the proton position in relation to the crystal ions as it feels a different electronic environment at different positions and its interaction with electrons changes.
$\gamma_{xy}({\bf R}_p)$ grows or decreases between extrema, whose locations correlate with the high-symmetry sites.
The amplitude of variations depend on the spatial directions and can be significant: e.g., in (a) and (b), $\gamma_{zz}$ varies by $40$ $\%$, while $\gamma_{xx}$ is nearly insensitive to the position; in (c), $\gamma_{xx}$ changes by a factor $3.5$ between the center of a face and the center of the cell.
Secondly, the matrix $\{\gamma_{xy}\}$ is generally anisotropic.
Differences in the diagonal elements along different crystallographic directions can be significant: in (d), $\gamma_{yy}$ and $\gamma_{zz}$ get up to  $1.7$ times larger than $\gamma_{xx}$; in (c), $\gamma_{xx}$ is up to $3.2$ times larger than $\gamma_{zz}$.
Thirdly, the friction tensor $\gamma_{xy}$ is generally non-diagonal: in (a) and (b), $\gamma_{xy}$ develops sizeable nonzero off-diagonal values at positions between the high-symmetry sites (about $30\%$ of $\gamma_{xx}$ at $x/a_c\!=\!0.25$).
The matrix is diagonal only at high-symmetry sites: in (c), when the proton is at the cube's center (corresponding to $x/a_c\!=\!0.5$) or in (a), when it is equidistant from the six nearest neighors ($x/a_c\!=\!0$ and $1$) .

\begin{figure}[t]
\includegraphics[scale=0.7]{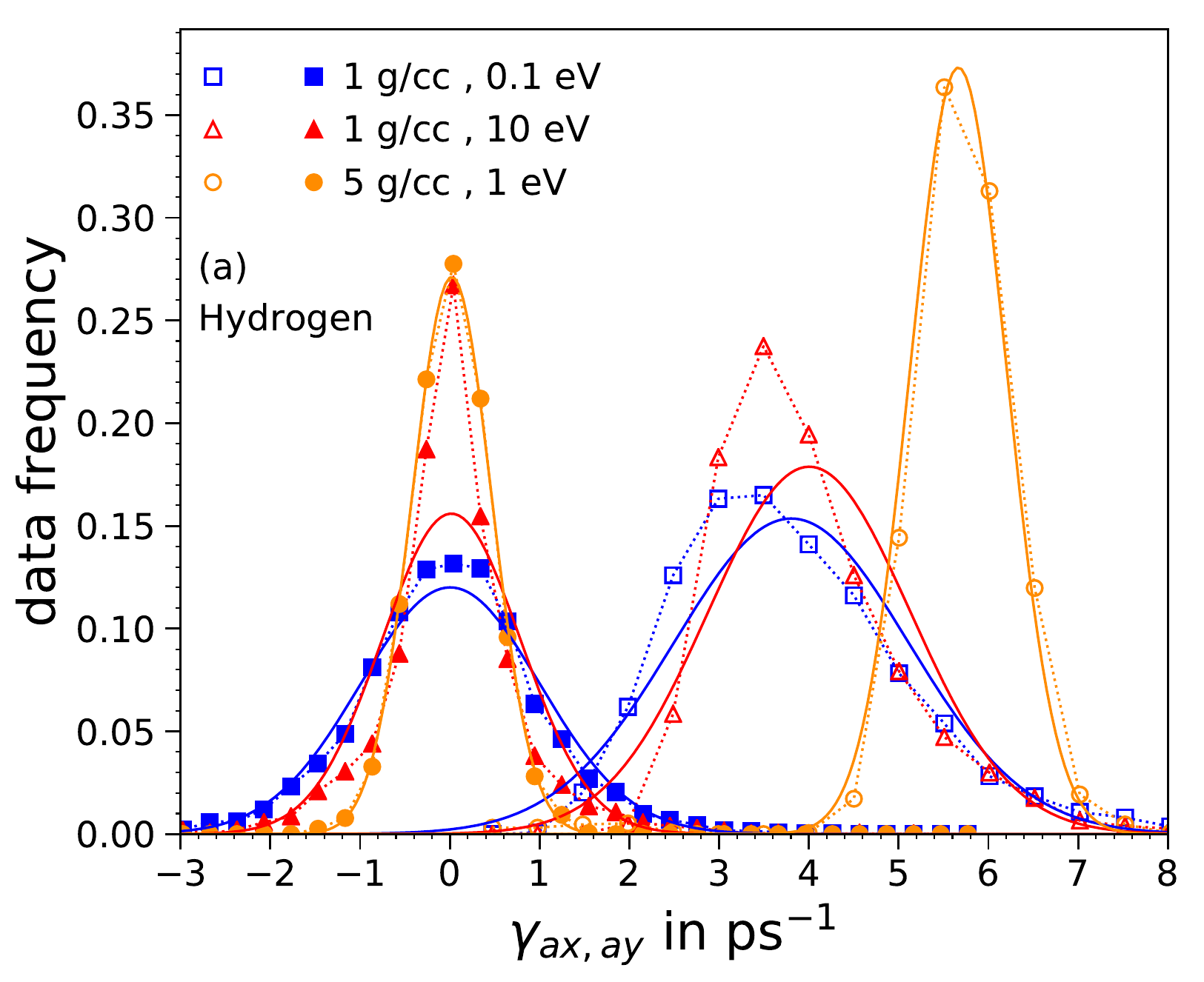}\\
\includegraphics[scale=0.7]{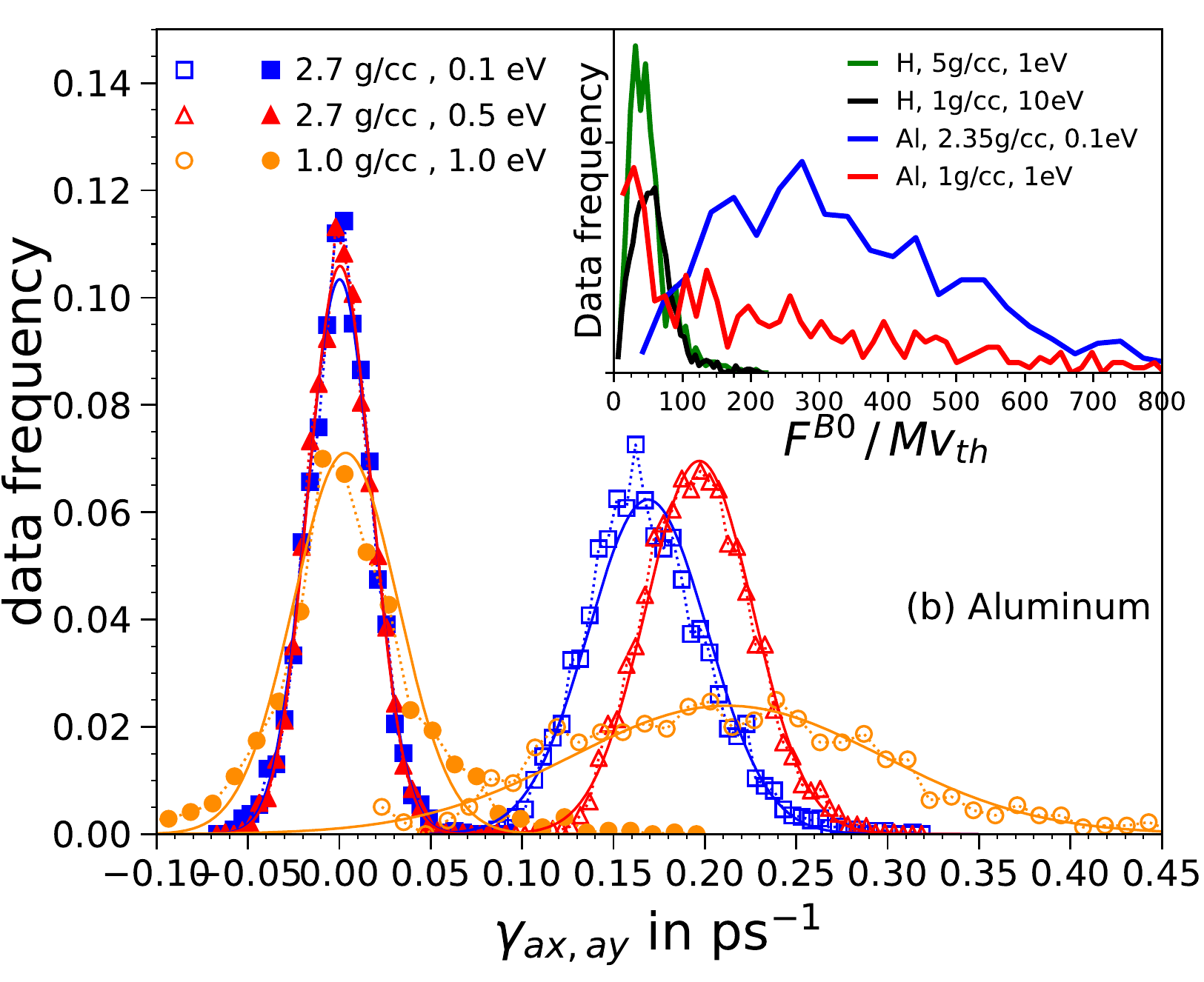}
\caption{(Color online) (a) Distributions of self diagonal (open symbols) and self off-diagonal (solid symbols) friction coefficients in warm dense H. The dashed lines show Gaussian functions calculated using the means and standard deviations of simulation data. (b) Same as in (a) for Al. The inset shows the histogram of BO forces in units of the mean friction force.}
\label{Fig:3}
 \vspace{-0.4cm}
\end{figure}
\begin{figure}[t]
\includegraphics[scale=0.7]{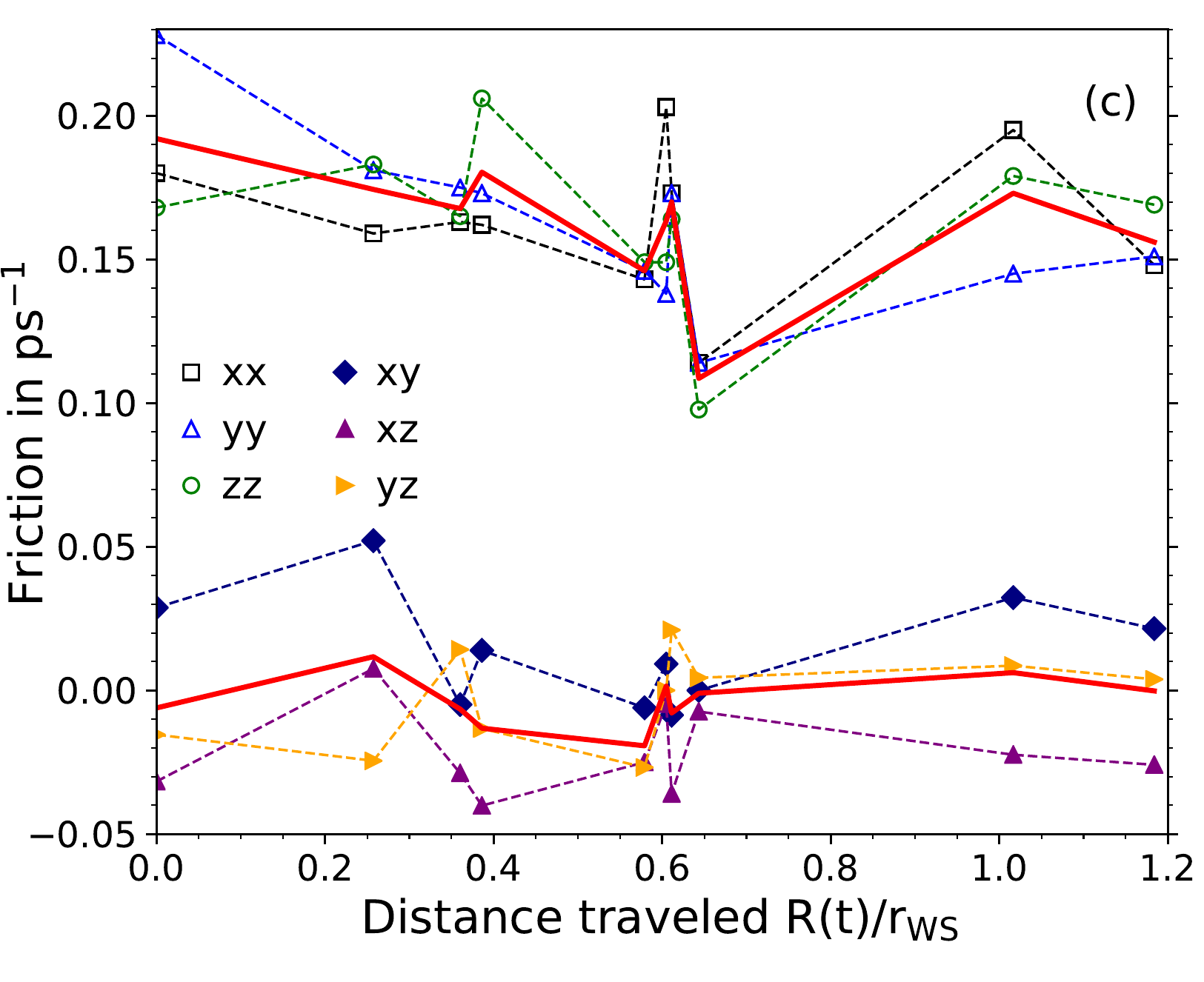}
\caption{(Color online) Self diagonal (open symbols) and self off-diagonal (solid) friction coefficients felt by an ion in Al at $2.35$ $\rm g.cm^{-3}$ and $0.1$ eV. The red lines show their mean values.}
\label{Fig:4}
\vspace{-0.4cm}
\end{figure}
\begin{table}[b]
\vspace{-0.2cm}
  \centering \begin{tabular}{|p{0.35cm}||p{0.55cm}|p{0.45cm}|p{0.72cm}|p{0.53cm}|p{0.55cm}||p{0.6cm}||p{0.6cm}|p{0.75cm}|p{0.45cm}|p{0.75cm}|}
    \hline
    
   &$ \rho$& $T$ &$\Theta$ & $\Gamma_{sc}$&$\kappa_{sc}$ & $\widetilde{\gamma}_d$ &${\gamma}_d$ & $\sigma_d$&${\gamma}_{od}$  &$\sigma_{od}$\\\hline
    &  1.0 & 0.1 & 0.004&   52.2& 1.3 & 4.57 &3.80 & 1.31& 0.0&1.0\\
    ${\rm H}$ & 1.0  & 10. & 0.39& 0.56& 1.25& 4.28 &4.0& 1.12& 0.0&0.77\\
    & 5.0&  1.0 & 0.013& 11.3& 1.1& 6.35 &5.66 & 0.54 & 0.0& 0.44\\\hline
    &  2.35 & 0.1 & 0.009 & 92 & 2.2& 0.21 &0.17 &0.033 &0.0&0.018\\    
    ${\rm Al}$ &  2.35 & 0.5 & 0.045& 18.5& 2.2& 0.22 &0.20& 0.029& 0.0&0.017\\
    &  1.0 & 1.0 & 0.17 & 5.3 & 2.4 & 0.48 & 0.21 & 0.085 & 0.0& 0.029\\\hline
  \end{tabular}
\vspace{-0.2cm}
\caption{Densities $\rho$ in $\rm g.cm^{-3}$ and temperatures $T$ in eV considered in this work. $\Theta\!=\!E_F/k_{\sss B}T$ with $E_F$ the Fermi energy. $\kappa_{sc}^{-1}$ is the finite-T Thomas-Fermi screening length in units of the average inter-ionic distance $r_{\sss WS}$, $\Gamma_s\!=\!(Ze)^2e^{-\kappa_{sc}}/r_{\sss WS}$ the effective coupling parameter of ions. The five rightmost columns are in $\rm ps^{-1}$.}
  \label{Tab:1}
\end{table}
We now analyze the electronic friction tensor in warm dense H and Al plasmas under the conditions listed in table~\ref{Tab:1}.
In each case, we have calculated all the elements of the tensor for $N_c\!=\!10$ ionic configurations equally spaced in time along a $5$ ps-long QMD simulation with $N\!=\!64$ or $128$ \cite{SM}.
Figures~\ref{Fig:3}a and \ref{Fig:3}b show the histograms of the self diagonal elements ($\gamma_{ax,ax}$ with $a\!=\!1$ to $N$ and $x\!=\!1,2,3$, open symbols) and of the self off-diagonal elements ($\gamma_{ax,ay}$ with $a\!=\!1$ to $N$ and $x,y\!=\!1,2,3$, $x\!\neq\! y$, solid symbols), including all the $N_c$ configurations.
As expected from isotropy and homogeneity, the distributions are independent of the Cartesian coordinate system and their mean values represent estimates of the canonical average discussed above.
In all cases, each quantity shows a single-peak distribution with mean and standard deviation given in table~\ref{Tab:1} (four rightmost columns).
For comparison, the full lines show the normalized Gaussian distributions obtained with these mean values.
For H, the distributions are nearly Gaussian at $5$ $\rm g.cm^{-1}$ but depart from a Gaussian law at $1$ $\rm g.cm^{-1}$ (e.g., the diagonal components have a right-skewed distribution).
For Al, they are all very nearly Gaussian. 

In all cases, we find that, unlike $\left\langle\tensor{\gamma}\right\rangle$, the self part of $\tensor{\gamma}$ is both anistropic and non-diagonal.
The diagonal components spread over a sizeable fraction of the mean value.
For instance, in Al, the full width at half maximum ($\simeq 2.355\sigma_d$) is $46\%$ of the mean value at $2.35$ $\rm g.cm^{-3}$ and $0.1$ eV and $34\%$ at $0.5$ eV, and it increases to $95\%$ at $1$ $\rm g.cm^{-3}$ and $1$ eV.
The increased dispersion at lower density is also seen in H, where, in addition, the tails of the skewed distributions at $1$ $\rm g.cm^{-3}$ extend to over two times the mean value.
The off-diagonal elements are typically much smaller than the diagonal components.
Yet, at lower density, they reach values comparable to the diagonal elements; e.g., in H at $1\rm g.cm^{-3}$ and $0.1$ eV the two distributions overlap.
The effect of the temperature is weak for the systems considered here.
For instance, $\gamma_d$ increases by $18\%$ between $0.1$ and $0.5$ eV in Al at $2.35$ $\rm g.cm^{-3}$, and by only $5\%$ between $0.1$ and $10$ eV in H at $1\rm g.cm^{-3}$.
These variations generally strongly depend on the details of the electronic DOS, and different behaviors can be expected in other metals \cite{Simoni_2019}.

To further understand the distributions, Fig.~\ref{Fig:4} shows the variations of $\gamma_{ax,ay}(R(t))$ for a randomly chosen Al ion $a$ as it travels through the Al plasma.
The coefficients are plotted versus the distance traveled from an initial position and measured at equally spaced time steps.
We find that, over the course of a quite short trajectory (the maximum distance traveled is $1.2$ times the average interionic distance $r_{\sss WS}$), the dispersion of the friction coefficients in Fig.~\ref{Fig:4} is similar to the dispersion of the distributions in Fig.~\ref{Fig:3}b (blue symbols).
Like in Fig.~\ref{Fig:2}, the variations correlate with the spatial variations of the electronic fluid along the ion trajectory, which, in the liquid-like state under consideration, consists in a succession of localized oscillations in the transient potential energy cages formed by neighbors followed by the passage into another cage \cite{Daligault_2006}.

We now make simple comparisons to assess the strength of frictional forces.
Firstly, it is interesting to compare the diffusion time scale $t_D\!=\!r_{\sss WS}^2/6D$, where $D$ is the self-diffusion coefficient, with the typical velocity relaxation time scale $t_\gamma\!=\!1/\gamma_d$.
For liquid density Al at $0.1$ eV, we find $t_D=0.46$ ps and $t_\gamma=5.9$ ps, i.e. $t_\gamma\!=\!\Delta R^2/6D$ with $\Delta R=3.6 r_{\sss WS}$\cite{Diffusion_Al}.
As expected, electronic friciton is weak, yet finite, corresponding to a theoretically difficult regime beyond the limits of standard methods (e.g., Smoluchowski equation \cite{Ermak1978}).
Secondly, the inset in Fig.~\ref{Fig:3}b shows the histogram of BO forces $||{\bf F}_a^{\sss BO}||/Mv_{th}$ measured for the same set of ionic configurations as in the main frames, in units of the mean frictional force on an ion with thermal velocity $v_{th}\!=\!\sqrt{k_{\sss B}T/M}$.
In H, the distributions are single peaked around $50$ at $1$ $\rm g.cm^{-3}$, $10$ eV and  around $35$ at $5$ $\rm g.cm^{-3}$, $1$ eV.
In Al, the distribution is peaked around $275$ at $2.35$ $\rm g.cm^{-3}$, $0.1$ eV and around $30$ at $1$ $\rm g.cm^{-3}$, $1$ eV.
Thus, albeit small, the frictional forces are not negligibly small compared to the BO forces.
As discussed in \cite{Simoni_2019, Daligault2019}, the small nonadiabatic couplings are responsible for the irreversible evolution toward equilibrium of the non-equilibrium states typically created in experiments \cite{Leguay2013,Cho_et_al_2015}, in the limit of thermal equilibrium they act as a thermostat to keep electrons and ions at the same temperature.
Although further work would be needed to assess their impact on the material properties, one cannot rule out at this stage that they are not strong enough, e.g., to affect the fluctuations that allow the potential barrier crossing events underlying particle diffusion and nucleation.

\begin{figure}[t]
\includegraphics[scale=0.7]{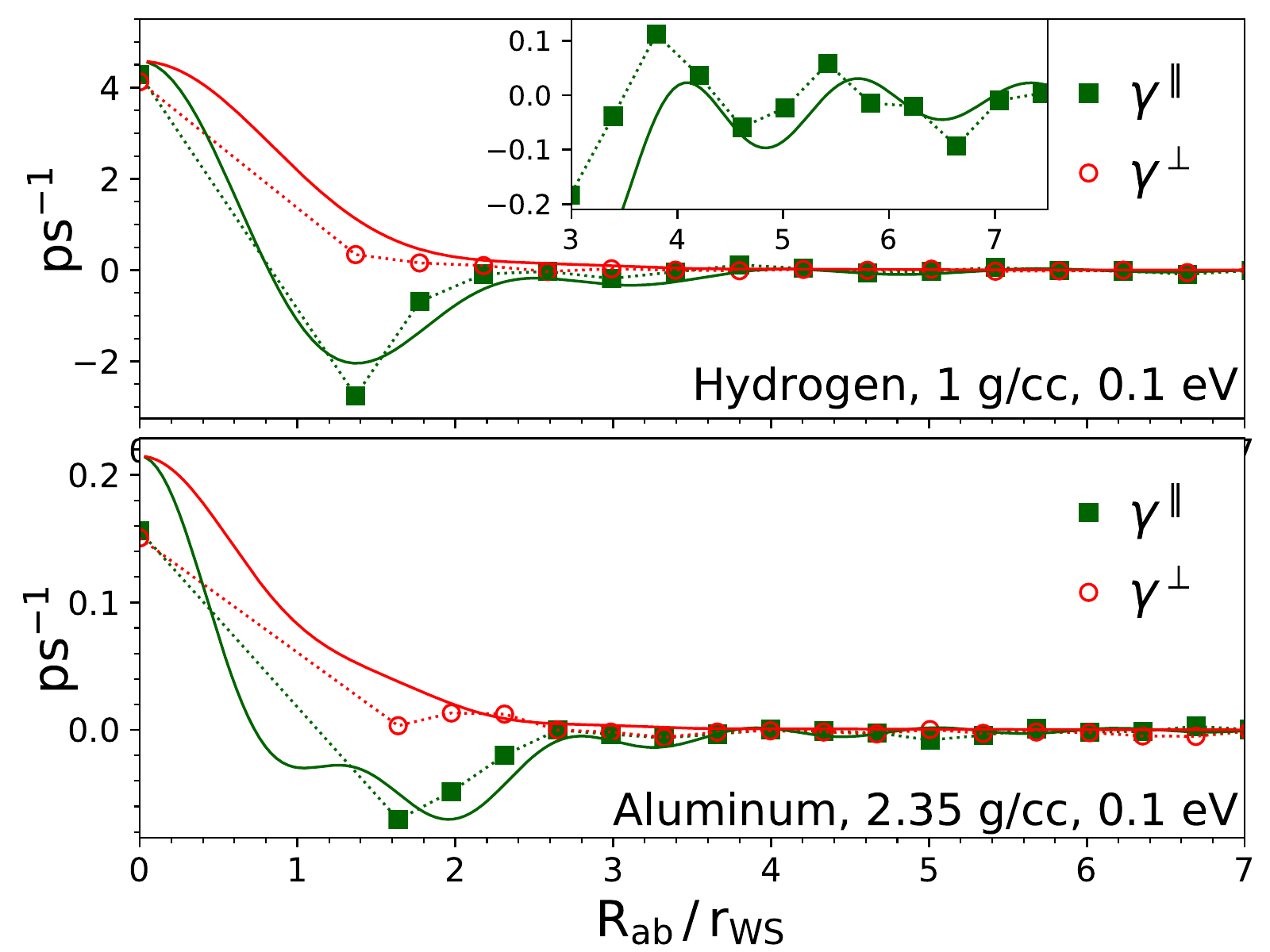}
\caption{(Color online) ``Parallel'' (squares) and ``perpendicular'' (circles) friction coefficients between pairs $(a,b)$ of ions vs their separation distance $R_{ab}$. The inset shows a zoom for $R_{ab}\!\ge\!3r_{\sss WS}$. The full lines show the predictions of the model.}
\label{Fig:5}
\vspace{-0.3cm}
\end{figure}
We now move on to discussing the ``cross'' terms $\gamma_{ax,by}$ that couple the motions of two distinct ions $a$ and $b$.
The detailed statistical analysis is challenging because each pair $(a,b)$ has a different orientation in the Cartesian coordinate system of the simulations and the distinction between diagonal (e.g., $xx$) and off-diagonal terms ($xy$) is meaningless.
We thus limit ourselves to an analysis of the ensemble averaged data.
For each pair $(a,b)$, we consider their coupling in a coordinate system where the x axis is directed along ${\bf R}_{ab}={\bf R}_a-{\bf R}_b$, and denote by $\gamma_{a,b}^{\parallel}$, $\gamma_{a,b}^{\perp_1}$ and $\gamma_{a,b}^{\perp_2}$ the diagonal elements in this coordinate system.
As expected by isotropy, upon averaging over all pairs of ions and over several ionic configurations, $\gamma_{a,b}^{\parallel}$ and $(\gamma_{a,b}^{\perp_1}+\gamma_{a,b}^{\perp_2})/2$ represent an estimate of $\gamma^\parallel$ and of $\gamma^\perp$.
We verified that upon averaging the latter depends only on the separation distance $R_{ab}$, that $\gamma_{a,b}^{\perp_1}$ and $\gamma_{a,b}^{\perp_2}$ become equal, and that the off-diagonal elements ($\parallel\perp_1$, etc.) vanish.
Figure~\ref{Fig:5} shows $\gamma^{\parallel}$ (solid symbols) and $\gamma^{\perp}$ (open symbols) as a function of $R_{ab}/r_{\sss WS}$ for H and Al systems; the value at $R_{ab}=0$ is set to the mean self diagonal friction ${\gamma}_d$ of Fig.~\ref{Fig:3}.
The data are compared to the predictions (full lines) of the model (\ref{tilde_gamma_axby}), which yields
\ben
\widetilde{\gamma}_{a,b}^{\parallel\/(\perp)}\!=\!-\frac{1}{M}\int_0^\infty{\!\!\! \frac{dk}{2\pi^2}k^4\! \left|\frac{v_{ei}(k)}{\epsilon(k)}\right|^2\!\frac{\partial{\rm Im}\chi_0(k,0)}{\partial\omega}f_{\parallel\/(\perp)}(kR_{ab})}\,,
\een
with $f_\parallel(r)\!=\!\frac{\sin(r)}{r}\!-\! 2 f_\perp(r)$ and $f_\perp(r)\!=\!\frac{\sin(r)}{r^3}-\frac{\cos(r)}{r^2}$.
As $R_{ab}$ increases, $\gamma^{\parallel}$ first reaches negative values at distances corresponding to the first layer of neighbors (the first peak of the pair-distribution function $g(r)$ (not shown) is at $r=1.6 r_{\sss WS}$).
The absolute magnitude of the first minimum is a significant fraction of ${\gamma}_d$ ($65 \%$ for H, $45 \%$ for Al).
The negative values mean that the total frictional force on $a$ is reduced (increased) when $a$ and $b$ move in the same (opposite) direction along ${\bf R}_{ab}$.
Beyond the first layer, $\gamma^{\parallel}$ slowly decays with $R_{ab}$ in an oscillatory manner around zero.
As for $\gamma^{\perp}$, it rapidly decays to values significantly smaller than ${\gamma}_d$.
Thus, the ``hydrodynamic'' couplings between ions mediated by electrons are mainly directed along the direction of separation, is sizeable between closest neighbors, and negligibly small with all the other ions.
Regarding the latter, one should nevertheless keep in mind the exact sum rule $\sum_{a}\sum_{x,y}\gamma_{ax,ay}=-\sum_{a\neq b}\sum_{x,y}\gamma_{ax,by}$ \cite{Daligault2019}, which couples all coefficients and results from momentum conservation.
The model remarkably reproduces these features even beyond the closest neighbors (see inset), which shows that the strength of the coupling by electronic friction is first a property of the electron gas that mediates it.

In summary, we have presented first-principle calculations of the electronic friction tensor $\tensor{\gamma}(R)$ in warm dense H and Al to characterize the frictional and random forces that affect the dynamics of ions in non-crystalline metallic systems due to their non-adiabatic interactions with electrons.
We have shown that, unlike the thermally averaged tensor and independently of the frame of reference, the instantaneous tensor is generally inhomogeneous, anisotropic and non-diagonal, and that these effects are stronger at lower density when electronic density variations are larger.
We have found that the nonadiabatic interactions introduce ``hydrodynamic'' coupling effects between the different ionic degrees of freedom, which is particularly sizable between nearest neighbors.
The model (\ref{tilde_gamma_axby}) gives a satisfactory description of the thermally averaged frictions and could be incorporated into classical molecular dynamics simulations.

\begin{acknowledgments}
This work was performed under the auspices of the U.S. Department of Energy under Contract No. 89233218CNA000001 and was supported in part by the U.S. Department of Energy LDRD program at Los Alamos National Laboratory through the grant No.20200074ER.
\end{acknowledgments}

\end{document}


\title{Supplemental Material for ``Nature of Non-Adiabatic Electron-Ion Forces in Liquid Metals''}
\author{Jacopo Simoni}\email[Contact email address: ]{jsimoni@lanl.gov}
\affiliation{Theoretical Division, Los Alamos National Laboratory, Los Alamos, NM 87545} 
\author{J\'erome Daligault} 
\affiliation{Theoretical Division, Los Alamos National Laboratory, Los Alamos, NM 87545}


\maketitle

\section{Details on Fig.~1 of the main paper}

The proton is displaced along the red and green line segments shown in the cartoon and defined as follows:\\
panel (a): FCC structure with proton position along the Miller direction $d=[\frac{1}{2}\frac{1}{2}0]$ starting at $R_o=(0,a_c/2,0)$.\\
panel (b): FFC structure, $d=[1 0 0]$, $R_o=(0,a_c/4,0)$.\\
panel (c): Simple cubic structure, $d=[1 0 0]$,  $R_o=(0,a_c/2,a_c/2)$ (dashed lines) and $R_o=(0,a_c/2,0)$ (full lines).\\
panel (d): BCC structure, $d=[1 0 0]$ with $R_o=(0,a_c/2,0)$.\\
When the components are equal by symmetry, only one is shown: in (a), $xx=yy$ and $xz=yz$; in (b), $xz=yz$; in (c), $yy=zz$ when $R_o=(0,a_c/2,a_c/2)$. In (c) and (d) all off-diagonal components are zero by symmetry.\\
The horizontal axis is the proton coordinate along $x$ in units of the cell size $a_c=7.65$ (FCC), $4.82$ (cubic) and $6.07$ (BCC) Bohr.

\section{The friction tensor calculation}

In order to compute the friction tensor $\gamma_{\rm ax,by}$ for the systems described in the main paper we combine a finite temperature Density Functional Theory (FT-DFT) approach that allows us to estimate the ground state of the electronic system at any given atomic configuration together with classical Molecular Dynamics (MD) simulations to temporally evolve the ionic positions. The ground state of the electronic system is obtained by solving the following set of Kohn-Sham (KS) equations.

\begin{equation}\label{Eq:KS}
  \hat{H}_\mathrm{KS}\ket{\Psi_{n,\mathbf{k}}} = \epsilon_n(\mathbf{k})\ket{\Psi_{n,\mathbf{k}}},
\end{equation}
where $\hat{H}_\mathrm{KS}$ is the KS Hamiltonian, $\ket{\Psi_{n,\mathbf{k}}}$ is the single particle Kohn-Sham state with energy eigenvalue $\epsilon_n(\mathbf{k})$. The KS Hamiltonian written on a spatial grid acquires the following form
\begin{equation}
  H_\mathrm{KS}(\mathbf{r}) = -\frac{\hbar^2\boldsymbol{\nabla}_\mathbf{r}^2}{2m_\mathrm{e}} + v_\mathrm{H}(\mathbf{r}) + v_\mathrm{xc}(\mathbf{r}) + v_\mathrm{ext}(\mathbf{r}),
\end{equation}
where the effective KS potential of the system is given by the sum of the Hartree component, $v_\mathrm{H}(\mathbf{r})$, the exchange-correlation component, $v_\mathrm{xc}(\mathbf{r})$, and the external ionic potential $v_\mathrm{ext}(\mathbf{r})$.

In order to obtain the friction tensor we numerically evaluate the following quantity
\begin{align}
\gamma_{\rm ax,by}(t) &= -\frac{1}{M}{\rm Re}\bigg<\sum_{n\neq m}\sum_{{\bf k}\in{\rm IBZ}}W_{\bf k}\frac{p_n({\bf k})-p_m({\bf k})}{\epsilon_n({\bf k})-\epsilon_m({\bf k})} f_{nm}^{\rm ax}({\bf k}) f_{mn}^{\rm by}({\bf k})\cos\bigg(\frac{\epsilon_n({\bf k})-\epsilon_m({\bf k})}{\hbar}t\bigg)\bigg>\,, \label{Eq:frict_coeff1} \\
\gamma_{\rm ax,by} &= \int_{0}^{\infty}dt\,\gamma_{\rm ax,by}(t)\,,\label{Eq:frict_coeff2}
\end{align}
that is formally equivalent to Eq.~(2) of the main paper, with the advantage to be easier to implement and compute numerically.
In Eq.~(\ref{Eq:frict_coeff1}) $<\ldots>$ is a thermal average over the ionic degrees of freedom, while the first summation is computed over all the possible transitions between the KS bands $n$ and $m$ with $n\neq m$ and the second over all the {\bf k}-points belonging to the Irreducible Brillouin Zone (IBZ). $p_n(\mathbf{k})=2(1 + e^{-\beta_\mathrm{e}(\mu-\epsilon_n(\mathbf{k}))})^{-1}$ is the Fermi-Dirac occupation for the spin unpolarized KS state $\ket{\Psi_{n,\mathbf{k}}}$ and $W_\mathbf{k}$ defines the {\bf k}-point integration weights.

The force matrix elements $f_{nm}^\mathrm{\rm ax}(\mathbf{k})$ associated to atom ${\rm a}$ are obtained from the following integral in real space
\begin{equation}\label{Eq:1}
  f_{nm}^\mathrm{\rm ax}(\mathbf{k}) = \hat{{\bf e}}_x\cdot\int_\Omega\,d{\bf r}\Psi_{n{\bf k}}(\mathbf{r})^* {\bf f}_{\rm a}({\bf r}) \Psi_{m\mathbf{k}}(\mathbf{r})\,,
\end{equation}
where $\Omega$ is the system's volume and ${\bf f}_{\rm a}({\bf r})$ is the effective electron-ion forces resulting from the electronic shielding of the bare electron-ion force centered on atom $a$. In all the calculations for both aluminum and hydrogen plasmas we employ local pseudo potentials.
\myFig{.7}{.7}{true}{0}{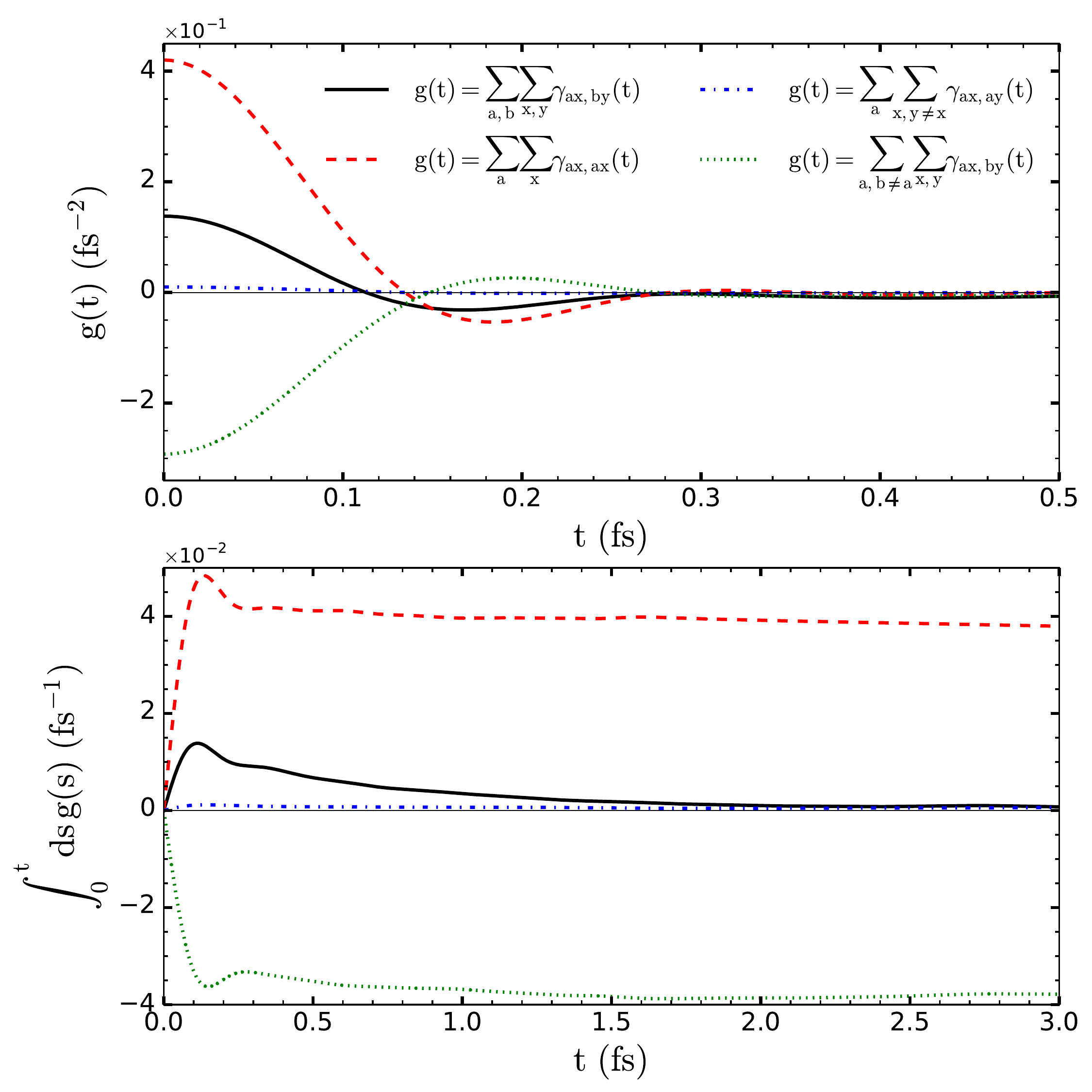}{(Color online) Analysis of different components of the tensor's time correlation function~(\ref{Eq:frict_coeff1}), upper panel, and of their cumulative sum~(\ref{Eq:frict_coeff2}), lower panel. The system considered is aluminum at $\rho=2.35\,g/cm^{3}$ and $T_{\rm i}=T_{\rm e}=0.5\,eV$.}{Fig:02}

\section{Details of the calculations}
All the QMD calculations presented in the main paper were performed by using the QUANTUM ESPRESSO $5.1$ program package\cite{QE2}, a typical calculation always consists of two main parts. The first part is a standard QMD simulation where the atoms evolve according to the Born-Oppenheimer dynamics. An Andersen thermostat is employed to ensure that the ionic temperature does not change during the temporal evolution, the system is first equilibrated and then let evolved for a sufficient amount of time (few picoseconds) allowing the accumulation of a number of well separated atomic configurations. The set of KS equations (\ref{Eq:KS}) are solved until convergence in the ground state electron density is reached. Then we can compute the forces acting on each ion and update the atomic positions at the successive MD step.

In the second part of the calculation we compute Eq.~(\ref{Eq:frict_coeff2}) for the friction tensor by averaging over several atomic configurations collected during the QMD run. For a given selected configuration a refined electronic structure calculation is performed where the number of bands, the energy cut-off $E_{\rm cut}$ of the plane wave expansion and the number of {\bf k}-points are increased in order to achieve full convergence. Table~(\ref{table1}) shows the set of chosen parameters in the case of different systems analyzed in the main paper (see Fig.~(2)).
\begin{table}
\caption{\label{table1} Typical parameters used during the self-consistent KS DFT calculations. The number of bands ($\mathcal{N}_b$), the number of {\bf k}-points ($\mathcal{N}_{\bf k}$), the cut-off energy ($E_{\rm cut}$), the number of atoms ($N_{\rm i}$) in the simulation box and the number of selected configurations ($N_{\rm c}$). }
\begin{ruledtabular}
\begin{tabular}{cccccccc}
&$T_{\rm e}\,\,(eV)$ & $\rho\,\,(g/cm^{3})$ & $\mathcal{N}_{b}$ & $\mathcal{N}_{\bf k}$ & $E_{\rm cut}\,\,(Ry)$ & $N_{\rm i}$ & $N_{\rm c}$\\
\hline
H & 0.1 & 1.0 & 100 & 64 & 150.0 & 128& 15\\
H & 1.0 & 5.0 & 400 & 64 & 150.0 & 128& 5 \\
H & 10.0& 1.0 & 1300& 8  & 150.0 & 128& 10\\
Al& 0.1 & 2.35& 250 & 8  & 150.0 & 64 & 18\\
Al& 0.5 & 2.35& 350 & 8  & 150.0 & 64 & 18\\
Al& 1.0 & 1.0 & 500 & 8  & 150.0 & 64 & 7 \\
\end{tabular}
\end{ruledtabular}
\end{table}
The number of bands $\mathcal{N}_{\rm b}$ used depends strongly on the electronic temperature of the system, the higher the electronic temperature is, the higher the number of bands required in the calculation is in order to converge. The number of atoms used in the simulation box is also an important parameter, in the case of aluminum plasmas we generally use $64$ atoms in the periodic box, while for hydrogen plasmas more atoms are usually necessary at a given temperature in order to generate a richer manifold of states and achieve a better convergence of the friction coefficients. For the same reason at low temperatures, in particular in the hydrogen case, a higher number of {\bf k}-points is also required.

We used in all the cases the Perdew-Zunger Local Density Approximation (LDA) to compute the exchange correlation potential, $v_\mathrm{xc}(\mathbf{r})$. However, the friction coefficients have a very weak dependence on the choice of the exchange-correlation functional.

\section{The exact sum rule}
From Fig.~(\ref{Fig:02}), lower panel, it is easy to observe that the exact sum rule derived in Ref.~\onlinecite{Daligault_2019}
\begin{equation} \label{Eq:G}
\lim_{t\rightarrow\infty}G(t) = \sum_{\rm ax,by}\gamma_{\rm ax,by} = 0\,,
\end{equation}
is satisfied by our tensor to a high degree of precision (see the black solid line in the figure). Here we show only the case of aluminum at liquid density and $T_{\rm i}=T_{\rm e}=0.5\,eV$, but the validity of Eq.~(\ref{Eq:G}) was verified in all the calculations presented in the main paper. The quantity $\sum_{\rm a}\sum_{x,y\neq x}\gamma_{ax,ay}$ is very low in magnitude compared to the other contributions to the tensor, this means that the validity of the sum rule is due to a perfect cancellation of the diagonal components of the tensor, $\sum_{\rm a}\sum_{x}\gamma_{ax,ax}$ (red dashed line), and of the remaining out of diagonal terms $\sum_{\rm a,b\neq a}\sum_{x,y}\gamma_{ax,by}$ (green dotted curve). This behaviour is common to all the other aluminum and hydrogen plasma cases that has been considered. 

